\documentclass[twocolumn,superscriptaddress,aps,showpacs,amsmath,footinbib,bibnotes]{revtex4-1}
\pdfoutput=1
\usepackage{graphicx}
\usepackage{amssymb,amsmath}
\usepackage{textcomp}
\usepackage[bold]{hhtensor}
\usepackage{natbib}
\usepackage{lipsum}
\usepackage{bm}
\usepackage{hyperref}
\usepackage{multirow}
\usepackage{float}
\usepackage{amsmath}
\usepackage{amssymb}
\usepackage{stmaryrd}
\usepackage{mathrsfs}
\usepackage[utf8]{inputenc}
\usepackage[export]{adjustbox}
\usepackage[dvipsnames]{xcolor}
\hypersetup{
	colorlinks,
	linkcolor={red!80!black},
	citecolor={Blue},
	urlcolor={MidnightBlue}
}

\begin{document}

\author{Amirhassan Shams-Ansari}
\thanks{These authors contributed equally to this work}
\affiliation{John A. Paulson School of Engineering and Applied Sciences, Harvard University, Cambridge, Massachusetts 02138, USA}
\author{Mengjie Yu}
\thanks{These authors contributed equally to this work}
\affiliation{John A. Paulson School of Engineering and Applied Sciences, Harvard University, Cambridge, Massachusetts 02138, USA}
\author{Zaijun Chen}
\thanks{These authors contributed equally to this work}
\affiliation{Max-Planck-Institut für Quantenoptik, Hans-Kopfermann-Str. 1, 85748 Garching, Germany }
\author{Christian Reimer}
\affiliation{John A. Paulson School of Engineering and Applied Sciences, Harvard University, Cambridge, Massachusetts 02138, USA}
\affiliation{HyperLight Corporation, 501 Massachusetts Ave, Cambridge, MA 02139, USA}
\author{Mian Zhang}
\affiliation{John A. Paulson School of Engineering and Applied Sciences, Harvard University, Cambridge, Massachusetts 02138, USA}
\affiliation{HyperLight Corporation, 501 Massachusetts Ave, Cambridge, MA 02139, USA}
\author{Nathalie Picqu\'{e}}\email{nathalie.picque@mpq.mpg.de}
\affiliation{Max-Planck-Institut für Quantenoptik, Hans-Kopfermann-Str. 1, 85748 Garching, Germany }
\author{Marko Lon\v{c}ar}\email{loncar@seas.harvard.edu}
\affiliation{John A. Paulson School of Engineering and Applied Sciences, Harvard University, Cambridge, Massachusetts 02138, USA}

\title{An integrated lithium-niobate electro-optic platform for spectrally tailored dual-comb spectroscopy}
\date{s\today}

\begin{abstract}
A high-resolution broad-spectral-bandwidth spectrometer on a chip would create new opportunities for gas-phase molecular fingerprinting, especially in environmental sensing. A resolution high enough to observe transitions at atmospheric pressure and the simultaneous sensitive detection of multiple atoms or molecules are the key challenges. Here, an electro-optic microring-based dual-comb interferometer, fabricated on a low-loss lithium-niobate-on-insulator nanophotonic platform, demonstrates significant progress towards such an achievement. Spectra spanning 1.6 THz (53 cm\textsuperscript{-1}) at a resolution of 10 GHz (0.33 cm\textsuperscript{-1}) are obtained in a single measurement without requiring frequency scanning or moving parts. The frequency agility of the system enables spectrally-tailored multiplexed sensing, which allows for interrogation of non-adjacent spectral regions, here separated by 6.6 THz (220 cm\textsuperscript{-1}), without compromising the signal-to-noise ratio.
\end{abstract}

\maketitle

The past decade has witnessed a remarkable progress in designs and technologies for chip-scale spectrometers relying on different spectrometric techniques based e.g. on interference \cite{pohl2020integrated,le2007wavelength,nie2017cmos,kita2018high,redding2013compact} or dispersion \cite{faraji2018compact}. However, acquiring a large number of spectral elements with a high resolution in a single measurement remains challenging. Recently, dual-comb spectroscopy combined with compact sources, such as fiber-doped mode-locked lasers or electro-optic (EO) modulators, has emerged as an intriguing approach \cite{picque2019frequency}. This technique measures the time-domain interference between two frequency combs of slightly different line spacings. The Fourier transform of the interference pattern reveals a radio-frequency (RF) spectrum made of the beat notes between pairs of comb lines, one from each comb. Similarly to Michelson-based Fourier transform spectroscopy \cite{griffiths2007fourier}, all the spectral elements are simultaneously measured on a single photo-detector, in a multiplexed fashion, resulting in spectra with an unparalleled consistency. The main distinguishing features of dual-comb spectrometers are the absence of moving parts, the use of coherent light sources that enhance the signal-to-noise ratio (SNR), the possibility of directly calibrating the frequency scale within the accuracy of an atomic clock, and narrow instrumental line shapes. Transferred to an integrated miniaturized device, such characteristics could significantly enhance the capabilities of in-situ real-time spectroscopic sensing. First proofs of concept towards an on-chip dual-comb spectrometer have been reported with Kerr combs \cite{suh2016microresonator, dutt2018chip,yu2018silicon,pavlov2017soliton}and quantum -and interband- cascade lasers \cite{villares2014dual,scalari2019chip}.Kerr combs rely on the third-order nonlinearity of the material and typically have a line spacing from hundreds of GHz to a few THz, which makes them more suited to condensed-matter spectroscopy \cite{dutt2018chip,yu2018silicon}. Although narrow-span Kerr combs can have line spacing of a few tens of GHz \cite{suh2016microresonator}, this is still too large for transitions of gas-phase species. Scanning the frequency of the comb lines via thermo-optic effect has been reported as a means to improve the resolution \cite{yu2018gas,lin2020broadband}. Electrically-pumped quantum- and interband-cascade lasers directly emit in the molecular fingerprint mid-infrared region, though their span and number of usable comb lines are small \cite{scalari2019chip,villares2014dual}. Here we explore a novel approach toward a spectrally-tailored on-chip spectrometer, based on frequency-agile frequency combs harnessing second-order nonlinearities \cite{zhang2019broadband}. We show that a low-loss integrated lithium niobate (LN) photonic platform \cite{zhang2017monolithic,luo2018highly}enables a significantly increased versatility for on-chip devices, providing tailored solutions to SNR optimization. We provide an experimental demonstration at a spectral resolution of 10 GHz, the highest resolution with a photonic-chip-based multiplexed or parallel-recording spectrometer so far.
Our EO dual-comb source consists of two racetrack optical resonators (on two separate chips), each with through and drop optical ports, integrated with a pair of microwave electrodes (Fig. 1) \cite{zhang2019broadband}. The resonators are formed by partially etching a 1.3-$\mu$m-wide waveguide into a 600-nm-thick X-cut LN device layer sitting on top of 2 $\mu$m of thermal SiO2. The fabricated devices are cladded with a 1-$\mu$m-thick SiO\textsubscript{2} layer (Fig. S1 - see Supplementary Information).The cross section of the waveguide is chosen to support a high Q-factor fundamental transverse electric mode (loaded Q ~10\textsuperscript{6}, Fig. S2). Furthermore, its weak normal dispersion enables broad-band frequency comb generation. For efficient EO interaction, the microwave electrodes are placed along the y-axis of the LN crystal in order to utilize the largest electro-optic coefficient (r\textsubscript{33}=30 pm/V). As compared to conventional EO-Comb sources \cite{jiang2007optical,carlson2018ultrafast} based on bulk LN crystals, the tight confinement of the light allows for electrodes to be placed close to the optical waveguide (~ 3.3 $\mu$m from each side) without introducing significant optical losses \cite{wang2018integrated}.
To generate an EO-Comb spectrum, each racetrack resonator is fed simultaneously with a continuous wave (CW) laser (frequency around 193 THz) and with a microwave synthesizer. The frequency of the microwave source is chosen to be the same as, or close to, the free-spectral-range of the racetrack resonator. In this way, three-wave mixing process is resonantly enhanced and results in an efficient sideband generation during each cavity roundtrip \cite{ho1993optical, kourogi1993wide,rueda2019resonant}.Owing to the low optical loss of the LN platform, the equivalent pathlength of the light is $(F/2\pi)\times L\sim 10 L$, where F is the cavity finesse, and L is the cavity roundtrip length (see Supplementary). The generated frequency comb beam is outcoupled via the drop port of the racetrack resonator, and collected using a lensed optical fiber. 

The multiplexed nature of Fourier transform spectroscopy provides an unparalleled consistency of the spectra, often at the expense of sensitivity \cite{griffiths2007fourier}: assuming that the total power onto the detector is kept constant, the SNR is inversely proportional to the number of spectral elements, here the comb lines. Detector nonlinearities usually set the maximum acceptable power to values much smaller than what is available from short-pulse laser sources. The widely-used solution to boost the SNR has been to use optical filters that select a single spectral band. With EO combs though, it becomes possible to simultaneously inject several CW lasers into each EO microring, labeled EO-Comb source1 and EO-Comb source2 in Fig. 1. Several (three in Fig. 1) pairs of mutually-coherent combs are thus produced. All combs generated in one microring share the same line spacing but their center frequencies are independently tunable. The detuning of the optical carriers and that of the microwave frequency are additional tuning knobs for adjusting the span and the shape of the spectrum \cite{zhang2019broadband}. This ability to tailor the spectra lifts the compromise between span and SNR: the interrogated regions are freely selected, and other domains may be left out, where e.g. there is no absorption or unwanted absorbing species exist, or where there is strong overlap between the transitions of the target and those of interfering species (such as water). The spectral tailoring can be dynamically changed and quickly adapted to new situations.

We first characterize a dual-comb set-up in a single spectral band using a single CW laser (Fig. S3) and no sample inserted in the beam path (see Supplementary). Since the two combs originate from the same CW laser, a good passive mutual coherence may be achieved, which is important for dual-comb interferometry. We measure the mutual coherence time of 5$\times$10\textsuperscript{-3} sec, currently limited by the fact that two comb sources reside on two different chips, located on two different optical benches. The lay-out of our dual-comb interferometer is similar to that used with bulk EO modulators \cite{long2014multiheterodyne,millot2016frequency}, where mutual-coherence times exceeding 1 s have been reported \cite{millot2016frequency}. Thus, we anticipate that similar performance will be possible in our approach, e.g. by fabricating the two microrings on the same photonic chip. The time-domain interference signal, the interferogram, is sliced in 5$\times$ 10\textsuperscript{-3}-s sequences, and a complex Fourier transform of each sequence provides amplitude and phase spectra with well-resolved individual comb lines. The averaged spectrum (95 second averaging time) reveals 160 comb lines (Fig. 2a), with a remarkable cardinal sine line shape (Fig. 2b) and a SNR culminating at 8$\times$ 10\textsuperscript{5} for the most intense comb line (Fig. 2c). The instrumental line shape (Fig. 2b), which perfectly follows the theoretical expectation of convolution of narrow beat notes by the Fourier transform of a boxcar, and the evolution of the SNR of the comb lines as the square-root of the averaging time (Fig. 2c) illustrate the high degree of interferometric control. The large tunability of the line spacing can be obtained by detuning the microwave signal driving the resonator, and it can be used to vary the refresh rate of the interferograms on demand (Fig. 2d). This could be of interest for broadband real-time dual-comb spectroscopy applications. However, similar to all other on-chip dual-comb spectrometers, the variation of the comb intensity profile currently limits the measurement time of our system.
A cell, filled with acetylene at close to atmospheric pressure, is then inserted and interrogated by EO-Comb source1, while EO-Comb source2 serves as the local oscillator. The transmittance (amplitude) and the dispersion (phase) in three spectra, averaged over 95 seconds each, are stitched to increase the span. The results are displayed in Fig. 3 for the spectral samples corresponding to maxima of comb lines. The resolution of 10.5 GHz, determined by the comb line spacing, is slightly larger than the self-broadened full-width at half-maximum of the rovibrational lines at 9.86$\times$ 10\textsuperscript{4} Pa, of 7.7 GHz, on average (see Supplementary). The residuals - the difference between the observed spectrum and HITRAN database \cite{gordon2017hitran2016}  – are below 10$\%$ with a standard deviation of 3.4$\%$. The two-fold improvement of the resolution and the acquisition of the dispersion spectrum, as compared to previously demonstrated integrated dual-comb spectrometers \cite{suh2016microresonator}, significantly improve the contrast to molecular absorption.
Finally, we demonstrate a first proof-of-concept of a spectrally-tailored dual-comb interferometer (Fig. 4). The LN electro-optic comb platform allows for the operation with multiple input lasers due to the inherent phase-locking mechanism established by the microwave driving, as well as the flexibility of the waveguide dispersion profile. In our experiment, we use two CW lasers, centered at \textit{f\textsubscript{1}=}192.7 THz and \textit{f\textsubscript{2}=}186.1 THz, to drive the EO-Comb source1, while the EO-Comb source2 is fed with acousto-optically frequency shifted light at frequency \textit{f\textsubscript{1}+$\delta$f\textsubscript{1}} and \textit{f\textsubscript{2}+$\delta$f\textsubscript{2}}, with \textit{$\delta$f\textsubscript{1}=}40 MHz and \textit{$\delta$f\textsubscript{2}=}25 MHz (Fig. 4). Two comb sources are driven with microwave signals of frequency \textit{f\textsubscript{RF1}=}10.453 GHz and \textit{f\textsubscript{RF2}=} \textit{f\textsubscript{RF1}}+0.1 MHz, respectively. As a result, two pairs of combs, each comprising 162 comb lines over a span of 1.7 THz, are generated with center frequencies that are 6.6 THz apart. Owing to the distinct acousto-optic frequency shifts, the two RF spectra do not overlap. The RF spectra could also be interleaved to achieve homogeneous SNR (see Supplementary) by choosing e.g. \textit{$\delta$f\textsubscript{2}=$\delta$f\textsubscript{1}+(f\textsubscript{RF2}-f\textsubscript{RF1})/2}, as long as the mutual coherence time of the system allows for resolved comb lines to disentangle the spectra. Even in this simple proof-of-principle demonstration, more than 100 spectral sections of 2000 comb lines each could be simultaneously measured, which represents overwhelming capabilities. The versatility of our integrated platform combined with the ability to probe transitions that are spectrally distant can open up opportunities to simultaneous detection of non-neighboring absorption lines that can belong to various molecules with optimized SNR. 
Compared to other on-chip demonstrations based on Kerr combs \cite{suh2016microresonator,dutt2018chip,yu2018silicon} or semi-conductor lasers \cite{villares2014dual,scalari2019chip}, our EO system points to a high versatility. The frequency agility and the spectral tailoring are its most promising features. Further improvement to the resonator’s Q-factor, while maintaining proper dispersion, could expand the spectral bandwidth of our sources towards an octave. This will enable broader interrogation regions for accessing many different species and self-referencing for high accuracy. LN’s wide transparency window (0.3-5 $\mu$m) can support EO comb sources both in the visible region \cite{desiatov2019ultra}, where electronic transitions of atoms and molecules are, and in the mid-infrared molecular fingerprint range. The availability of various components in our LN platform, such as acousto-optic modulators \cite{cai2019acousto} and phase-modulators \cite{ren2019integrated}, can pave the way towards a fully integrated photonic circuit for the dual-comb spectrometer. Although the resolution reported in this work is the highest so far for a fully-multiplexed photonic chip, new strategies for reaching the requirements of gas-phase spectroscopy are needed and will be implemented in the near future. In addition, shared with all existing on-chip systems, the decrease in the intensity of the comb lines from the carrier frequency narrows the spectral span and increases the measurement times. Overcoming the limitations of the dynamic range of on-chip comb sources posts another challenge to device fabrication and photonic design. Nevertheless, the thin-film lithium-niobate platform, with its unique combination of ultra-low losses and strong second- and third-order nonlinearities, promises an incomparable set of novel opportunities for on-chip high-resolution spectrometers.
\begin{acknowledgments}
We thank Dr. Edward Ackerman (Photonicsystems) and Dr. Yoshitomo Okawachi (Columbia University) for help with the experiment. Funding: This work is supported in part by Air Force Office of Scientific Research (AFOSR; award number of FA9550-19-1-0310), National Science Foundation (NSF PFI-TT; award number of IIP-1827720), Defense Advanced Projects Agency (DARPA; W31P4Q-15-1-0013) and by the Max-Planck Society; Device Fabrication is performed at the Harvard University Center for Nanoscale Systems, a member of National Nanotechnology Coordinated Infrastructure Network, which is supported by the National Science Foundation (award number of ECCS-1541959). 
\end{acknowledgments}
\bibliographystyle{ieeetr}
\bibliography{bibliography}
  \begin{figure*}
	\centering
	\includegraphics[width=0.8\linewidth,height=\textheight,keepaspectratio]{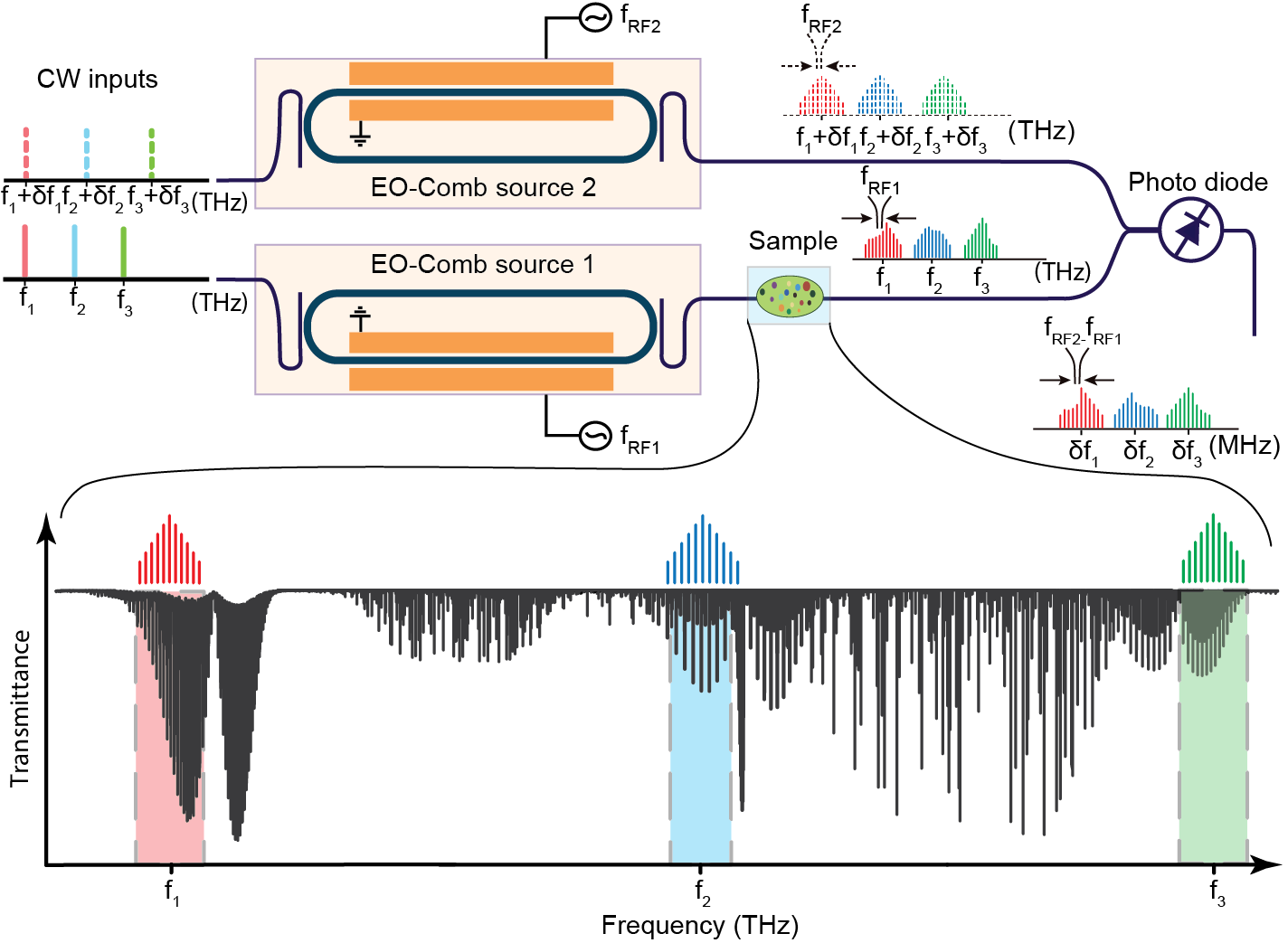}

	\caption{\label{fig1}\textbf{Spectrally-tailored dual-comb spectroscopy with microring electro-optic (EO) frequency combs. }\textbf{(a)} 
	In this conceptual representation, three narrow laser lines of frequency \textit{f\textsubscript{1}}, \textit{f\textsubscript{2}}, and \textit{f\textsubscript{3}} are used to excite the optical modes of a resonant EO modulator, driven at the microwave frequency \textit{f\textsubscript{RF1}} that corresponds to the resonator free-spectral range. The generated composite comb contains three non-adjacent combs, centered at \textit{f\textsubscript{1}}, \textit{f\textsubscript{2}}, and \textit{f\textsubscript{3}} each having comb-line spacing of \textit{f\textsubscript{RF1}}. Central frequency and span of each of the three combs are chosen to interrogate targeted features in an absorbing sample, while leaving out regions without absorber or with interfering species, and can be adjusted independently with agility. Three slightly frequency-shifted replica of the original laser lines are produced and injected into a second resonant EO modulator, driven at a slightly different repetition frequency \textit{f\textsubscript{RF2}}, providing a reference spectrum. The interrogating and reference comb beams interfere on a fast photodetector. In this process, pairs of comb lines, one from each beam, produce a composite radio-frequency (RF) comb of line spacing \textit{f\textsubscript{RF2}-f\textsubscript{RF1}}, mapping the spectral information from the optical domain, centered at \textit{f\textsubscript{1}}, \textit{f\textsubscript{2}}, and \textit{f\textsubscript{3}} to the RF domain, centered at \textit{$\delta$f\textsubscript{1}}, \textit{$\delta$f\textsubscript{2}}, \textit{$\delta$f\textsubscript{3}}, respectively. The frequency agility of the resonant EO modulator is a unique feature that enables on-demand spectral tailoring and optimization of the signal-to-noise ratio.}
\end{figure*}

  \begin{figure*}
	\centering
	\includegraphics[width=0.8\linewidth,height=\textheight,keepaspectratio]{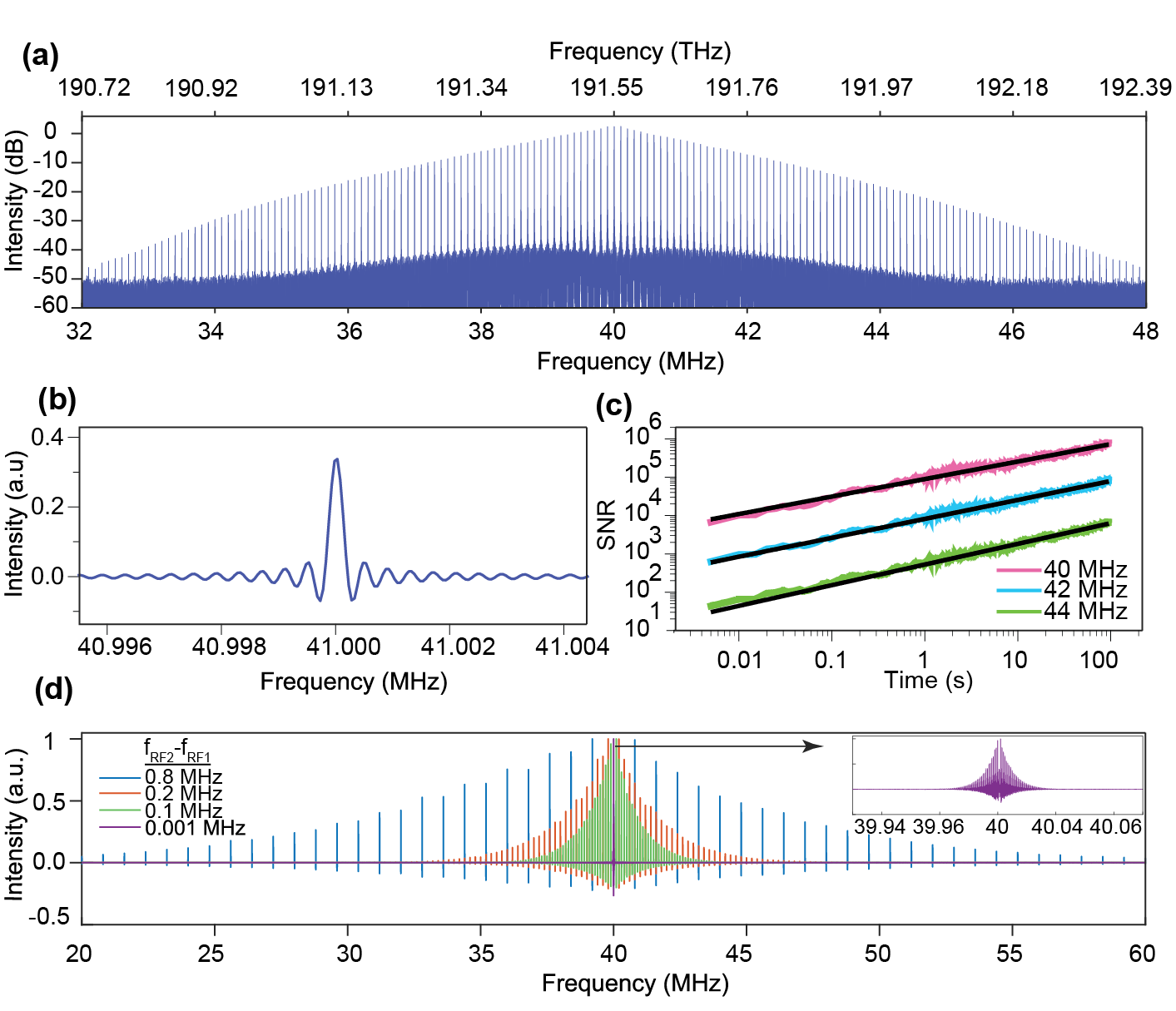}

	\caption{\label{fig2}\textbf{Experimental results of the microring electro-optic (EO) dual-comb spectrometer.}\textbf{(a)} 
	Apodized dual-comb spectrum (\textit{f\textsubscript{1}=}191.55 THz and \textit{$\delta$f\textsubscript{rep1}=} 40 MHz) with the measurement time of 95 seconds. 19000 interferograms (5 ms each) are acquired and processed, and the amplitude spectra are averaged. The two EO combs are driven with microwave frequencies of 10.0000 GHz and 10.0001 GHz (\textit{f\textsubscript{RF2}-f\textsubscript{RF1}=} 0.1 MHz), respectively. The signal-to-noise ratio (SNR) of the center comb-line is 8$\times$10\textsuperscript{5}, and the average SNR of 160 comb lines is 1$\times$10\textsuperscript{5}. \textbf{(b)} An unapodized individual comb line, near 41 MHz in (a), is shown featuring a cardinal-sine instrumental line shape. \textbf{(c)} The evolution of the SNR of three selected comb lines from (a) at 40 MHz, 42 MHz, and 44 MHz over the measurement time. Their linear fitted slopes (black line) are 0.46, 0.49, and 0.54, respectively, indicating that SNR increases with the square root of the measurement time.\textbf{(d)} The reconfigurability of the microring EO dual-comb system. The interferogram refresh rate \textit{f\textsubscript{RF2}-f\textsubscript{RF1}} is varied on demand at 0.800, 0.200, 0.100, and 0.001 MHz.}
\end{figure*}

 \begin{figure*}
	\centering
	\includegraphics[width=0.8\linewidth,height=\textheight,keepaspectratio]{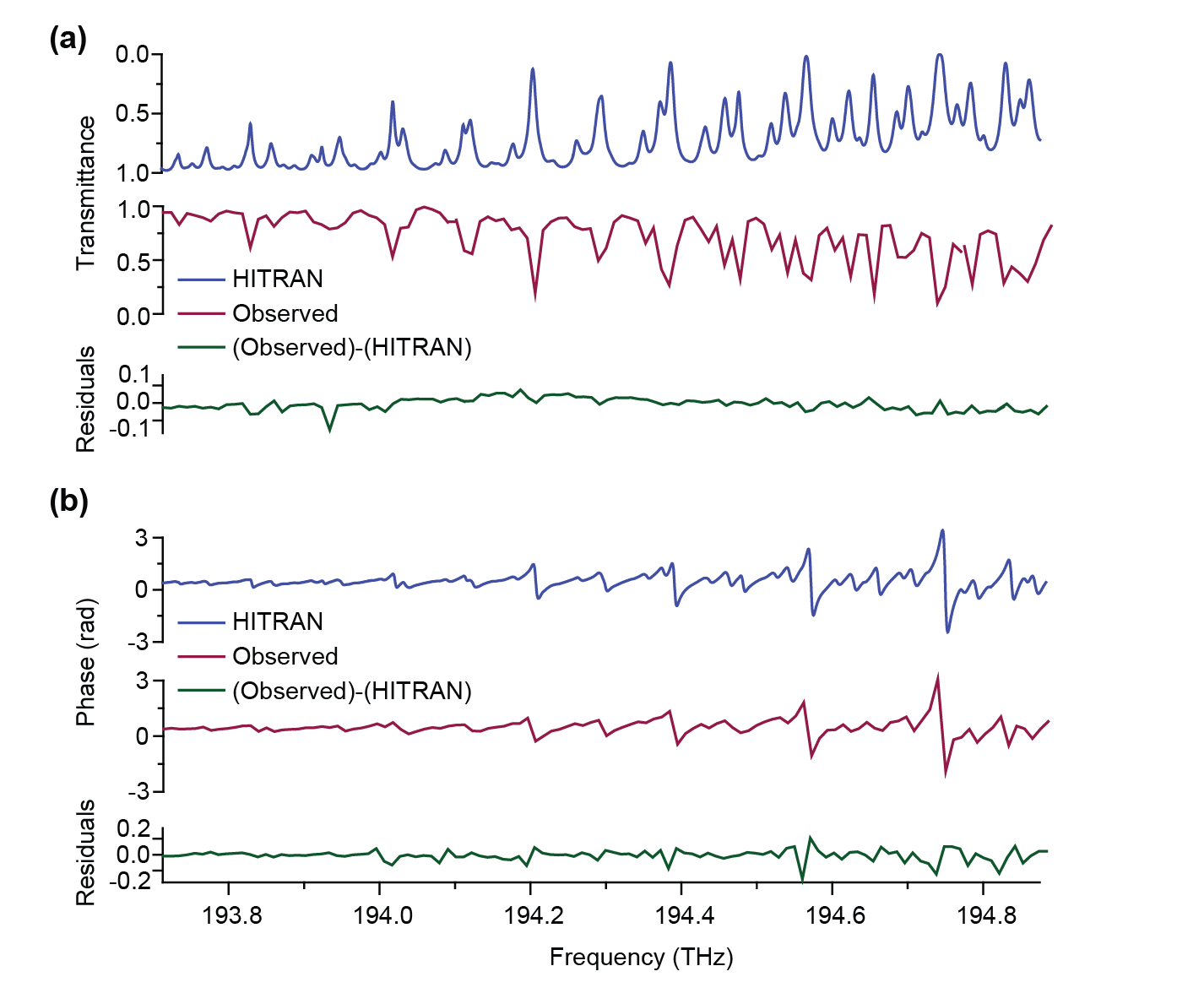}

	\caption{\label{fig3}\textbf{Measured dual-comb absorption and dispersion spectra from the P(34) to the P(22) lines of the $\nu$\textsubscript{1}+$\nu$\textsubscript{3} band of \textsuperscript{12}C\textsubscript{2}H\textsubscript{2}}\textbf{(a)} A multipass cell with 80 cm of absorption length, filled with acetylene at a pressure of 9.86$\times$10\textsuperscript{4} Pa, is interrogated \textbf{(a)} Experimental transmittance spectrum (red) and transmittance spectrum (blue) computed from the line parameters available in the HITRAN database. The spectral resolution is 10.453 GHz. Three spectra, centered at frequencies of 193.93 THz, 194.56 THz and 194.64 THz, are stitched. The residuals (green) - the difference between the observed spectrum and HITRAN – are below about 10$\%$ with a standard deviation of 3.4$\%$. \textbf{(b)} Experimental dispersion spectrum (red) and dispersion spectrum simulated from HITRAN database (blue). The residuals (green) are below 10$\%$ with a standard deviation of 2$\%$.}
\end{figure*}

 \begin{figure*}
	\centering
	\includegraphics[width=0.8\linewidth,height=\textheight,keepaspectratio]{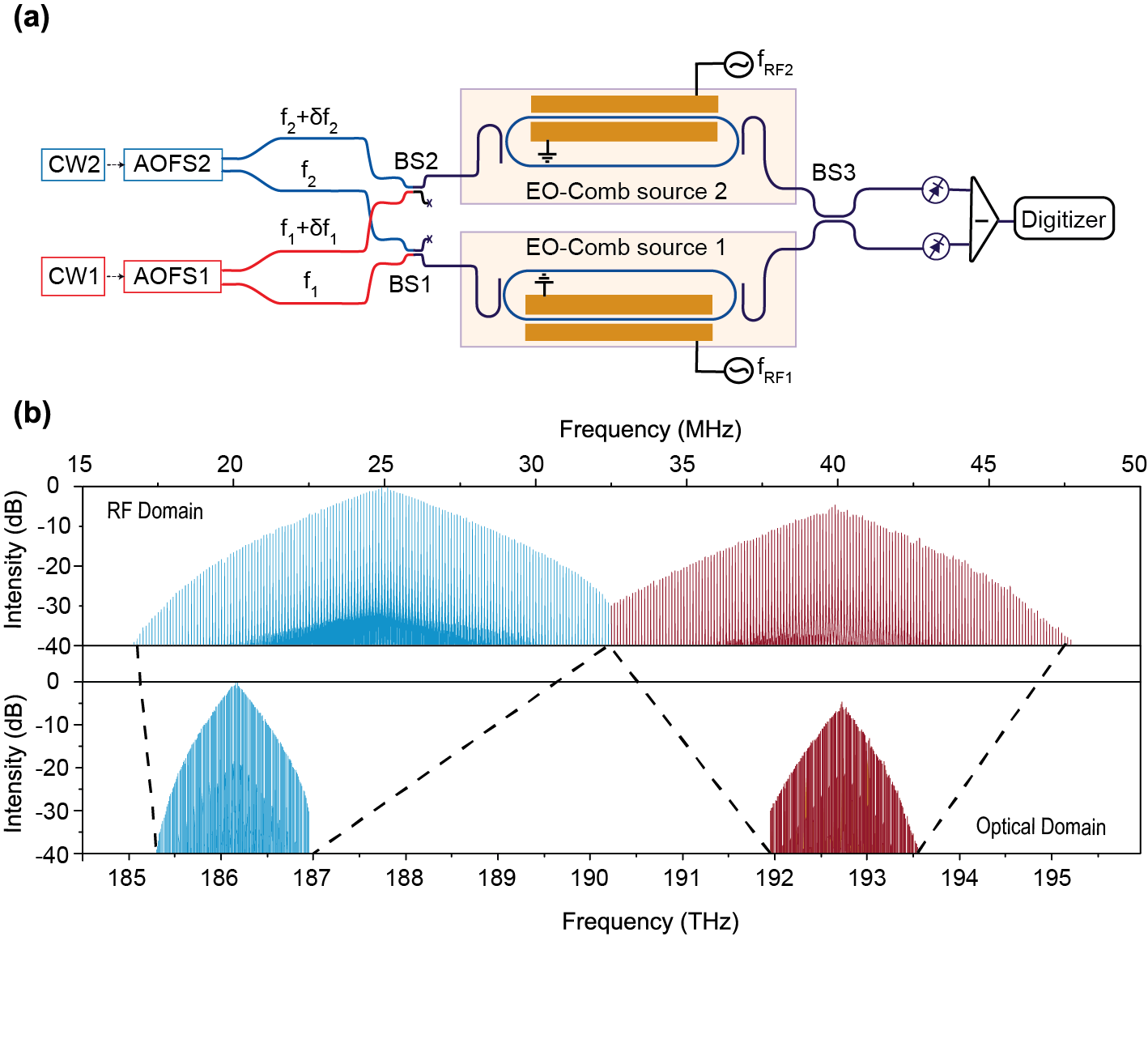}

	\caption{\label{fig4}\textbf{Spectrally-tailored dual-comb spectrometer.}\textbf{(a)}Experimental set up. CW, continuous-wave laser; AOFS, acousto-optic frequency shifter; BS, beam-splitter. CW1 and CW2 with \textit{f\textsubscript{1}=} 192.7 THz and \textit{f\textsubscript{2}=}186.1 THz pass through two AOFS (with $\delta$\textit{f\textsubscript{1}=}40 MHz, and $\delta$\textit{f\textsubscript{2}=}25 MHz), respectively. CW1 and CW2 are injected into EO-Comb source1 (\textit{f\textsubscript{RF1}=}10.4530 GHz) while their frequency-shifted replica are sent to EO-Comb source2 (\textit{f\textsubscript{RF2}=}10.4531) GHz with \textit{f\textsubscript{RF2}-f\textsubscript{RF1}=}0.1 MHz. The outputs are heterodyned on a balanced detector, and the time-domain interference signal is digitized with a data-acquisition board. \textbf{(b)} Dual-comb spectrum with a measurement time of 9.8 seconds. The two RF combs, which center frequencies are 15 MHz apart, correspond to two EO combs that are 6.6 THz apart in the optical domain.}
\end{figure*}

\end{document}